\documentclass[prb,showpacs,twocolumn]{revtex4}

\usepackage{epsfig}
\usepackage{dsfont}
\usepackage{amsmath}

\begin{document}
\bibliographystyle{revtex}

\title{Dynamics of quantum coherences at strong coupling to a heat bath}

\author{Denis Kast and Joachim Ankerhold}
\affiliation{Institut f\"ur Theoretische Physik, Universit\"at Ulm, Albert-Einstein-Allee 11, 89069 Ulm, Germany}

\date{\today}

\begin{abstract}
  The standard approach for path integral Monte Carlo simulations of open quantum systems is extended as an efficient tool to monitor the time evolution of coherences (off-diagonal elements of the reduced density matrix) also for strong coupling to environments.
  Specific simulations are performed for two level systems embedded in Ohmic and sub-Ohmic reservoirs in the domains of coherent and incoherent dynamics of the polarization. In the latter regime, the notorious difficulty to access the long time regime is overcome by
    combining simulations on moderate time scales with iteratively calculated initial densities.
    This allows to extract relaxation rates for sub-Ohmic environments at finite temperatures and over a broad range of couplings and to compare them to analytical predictions. The time evolution of the von Neumann entropy provides insight into the quantum phase transition at thermal equilibrium from a delocalized to a localized state at zero temperature.
\end{abstract}


\maketitle

\section{Introduction}

The dynamics of open quantum systems has again received considerable interest in the last years \cite{weiss,breuer}, mainly due to the progress in fabricating and manipulating quantum devices with unprecedented accuracy. Examples include complex superconducting circuits and tailored set-ups for ensembles of ultra-cold atomic gases. In fact, the interaction to reservoirs has even been found as a resource to drive the time evolution of complex system into desired final states \cite{zoller}.

Accordingly, there have been substantial efforts to develop new and to extend existing numerical methodologies to treat open system dynamics beyond conventional perturbative approaches like e.g. master equations. Among them are those that work with the reduced density operator of the relevant system and those that treat the full Hilbert space of system + reservoir. The latter comprise renormalization group methods \cite{nrg} (NRG/DMRG) and advanced basis set techniques \cite{mctdh} (MCTDH), the former ones include
 the quasi-adiabatic propagator approach \cite{quapi} (QUAPI), stochastic Schr\"odinger \cite{lehur} and Liouville-von Neumann  \cite{stockburger} (SLN) formulations, and path integral Monte Carlo schmemes (PIMC)\cite{egger,muehlbacher1}. Each of these methods has its strengths and its limitations. The most challenging situations appear in the domains of strong coupling to heat baths with non-Ohmic spectral densities at low but elevated temperatures. There, the non-Markovian retardation is influential and system-bath correlations are affected by both quantum and thermal fluctuations.

The PIMC approach belongs to the very few methods that yield numerically exact results for the
dynamics in all ranges of parameter space, particularly, in the deep non-perturbative regime. It is based on the formally exact Feynman-Vernon formulation of the reduced dynamics in terms of path integrals \cite{weiss}. The impact of the reservoir is captured by a so-called influence functional which contains the force-force correlation of the bath and thus carries a retarded interaction in time between system paths. Impressive progress has been achieved in recent years to extend it to a very efficient and powerful tool for the quantum dynamics of systems with discrete Hilbert space (tight-binding systems) \cite{muehlbacher1}. Single and correlated charge transfer along molecular chains\cite{muehlbacher1,muehlbacher2},
the impact of external driving fields \cite{muehlbacher3} and steady states in charge transfer through quantum dots and molecular contacts \cite{muehlbacher4} have been studied.

While in these applications the reservoirs are basically Ohmic (typically with a large cut-off), recently sub-Ohmic reservoirs and their entanglement with the system of interest gained attention \cite{subohmic1,subohmic2,subohmic3,subohmic4}. These reservoirs have spectral densities of the form $J(\omega)\sim \alpha\, \omega^s$ with typical coupling constant $\alpha$ and spectral exponent $s$.
They are not only generic models to better understand fundamental concepts in quantum mechanics, but also to explain at least qualitatively experimental observations for superconducting qubits \cite{fnoise} and quantum dots \cite{ring}, quantum impurity systems \cite{heavyf}, nanomechanical oscillators \cite{nanomech}, and trapped ion systems \cite{marquardt}. In this context, a generic model is that of a two level system (artificial spin-$\frac{1}{2}$) interacting with a heat bath (spin-boson model\cite{leggett,weiss}, SBM). The SBM has been analyzed in variety of arrangements in the past\cite{weiss}, however, its sub-Ohmic properties have been left basically untouched mainly due to the lack of powerful numerical schemes. In the last years, this gap has been closed by applying advanced techniques mentioned above. It turns out that the dominance of low frequency modes causes strong retardation effects in the dynamics, quantum phase transitions in thermal equilibrium at zero temperature\cite{subohmic1} accompanied by  substantial system-reservoir entanglement\cite{hofstetter}.
Very recently, the domain of strong friction, unaccessible by alternative approaches, has been investigated with PIMC simulations and revealed the persistence of coherences\cite{kast}.

For that purpose, the approach has to be extended to measure not only polarizations (populations), but also off-diagonal elements of the density. In this work we go one step further and use the knowledge of the full reduced density matrix to improve the PIMC technique to cover also long times. Namely, due to the dynamical sign problem, i.e.\ an exponentially decreasing signal to noise ratio with growing simulation time, for the PIMC method are not reachable using the PIMC method. The idea is now to glue together sequences of PIMC simulations for moderate times with successively determined initial densities. In general,
this procedure does not work due to intricate system-bath correlations. For moderate to strong friction though, coherences in the system dynamics are strongly suppressed even at very low temperatures and the relaxation towards thermal equilibrium is monotone. Corresponding rates are still quantum mechanical and can be extracted efficiently from relatively cheap long time simulations based on this concept.

The paper is organized as follows: The model and relevant observables are introduced in Sec.~\ref{model} to lay the basis for a brief description of the PIMC in Sec.~\ref{pimc} and analytical treatments in Sec.~\ref{niba}. The implementation of the measurement operator for the coherences and the extended PIMC is first studied in Sec.~\ref{ohmic} for the much simpler Ohmic case as a test-bed. In Sec.~\ref{subohmic} the challenging sub-Ohmic SBM is considered, particularly its relaxation rates in the incoherent domain and the entropy. Conclusions are given in the end.

\section{Model and observables}\label{model}
The spin-boson model (SBM) describes a two-level system (TLS) linearly coupled to a heat bath environment. This is modeled as an ensemble of harmonic oscillators according to the rationale that reservoirs with macroscopic many degrees of freedom carry Gaussian fluctuation properties. Here, we focus on the symmetric situation
\begin{eqnarray} \label{hamilton}
H &=& H_{TLS} + H_I + H_B   \nonumber \\
&=& -\frac{\hbar\Delta}{2}\sigma_x
    - \frac{\hbar \sigma_z}{2} \sum_{\alpha} \lambda_{\alpha} (b^{\dag}_{\alpha} + b_{\alpha})
    + \sum_\alpha \hbar \omega_\alpha b^{\dag}_{\alpha} b_{\alpha}\, \nonumber\\
\end{eqnarray}
with degenerate eigenstates $|\pm 1\rangle$ of the Pauli $\sigma_z$ operator coupled by a tunneling energy $\hbar\Delta$. In the continuum limit, the distribution of the bath frequencies is described by a spectral density $J(\omega)=\pi \sum_\alpha  \lambda_\alpha^2\, \delta(\omega_\alpha-\omega)$.

We are interested in the time evolution of the observables
\begin{eqnarray}\label{observables}
P_\nu(t) & = & \langle \sigma_\nu(t) \rangle  \nonumber \\
       & = & {\rm Tr_S} \left\{\sigma_\nu \, \rho(t) \right\}\ , \ \nu=x, y, z
\end{eqnarray}
which fully characterize the reduced density
\begin{equation}\label{reduced}
\rho(t)={\rm Tr}_B\left\{\exp(-iHt/\hbar) W(0) \exp(iHt/\hbar)\right\}\, .
\end{equation}
The initial state is assumed to be a product state of the form
\begin{equation}\label{prep}
W(0) = \frac{1}{Z_B} {\rm e}^{-\beta  (H_B-\mu{\cal E})} \rho(0)
\end{equation}
with the collective bath operator ${\cal E} = \sum_{\alpha} \hbar\lambda_{\alpha} (b^{\dag}_{\alpha} + b_{\alpha})$, the reservoir's partition function $Z_B$, and the initial state $\rho_0$ of the bare TLS. The polarization parameter $\mu$ allows to start with shifted initial bath distributions. For example, for $\mu=\pm 1$ the reservoir is equilibrated to an initial spin state $|\pm 1\rangle$ (polarized), while for $\mu=0$ one regains  the bare heat bath.

In contrast to previous PIMC simulations, we consider for the initial state $\rho_0$ not only eigenstates of the $\sigma_z$-operator (localized spin states), but also generalized initial preparations
\begin{equation}\label{eq:initialprep}
\rho(0) = \frac{1}{2} \left(\mathds{1} + \sum_{\nu=x,y,z} P_\nu(0)\,  \sigma_\nu \right)\,
\end{equation}
with properly chosen initial values $P_\nu(0)$ fulfilling the constraint $\sum_\nu P_\nu(0)^2=1$.

\section{Path Integral Monte Carlo Technique}\label{pimc}
A non-perturbative treatment of the dynamics (\ref{observables}) is obtained within the path integral formulation.
The $P_\nu$ is expressed along a Keldysh contour with forward
$\sigma$ and backward $\sigma'$ paths \cite{weiss}. The impact of the environment appears as an
influence functional introducing arbitrarily long-ranged
interactions between the paths. Switching to sum and difference paths, respectively,
\begin{equation}
\eta=\frac{\sigma+\sigma'}{2}\ \ , \  \ \zeta=\frac{\sigma-\sigma'}{2}\, ,
\end{equation}
one
arrives at
\begin{equation}\label{pathinte}
P_\nu(t)=\int {\cal{D}}[\eta]{\cal{D}}[\zeta]{\cal{A}}_\nu\, {\rm e}^{-{\Phi}[\eta,\zeta]}
\end{equation}
with the contribution $\cal{A}_\nu$ in absence of dissipation including the measurement operator and the influence functional
\begin{eqnarray}
   \Phi[\zeta, \eta] &=& \int_0^t dt' \int_0^{t'} dt'' \zeta(t') \left[ \ddot{Q}'(t'-t'') \zeta(t'') \right. \nonumber \\
   & +&\left. i \ddot{Q}''(t'-t'') \eta(t'') \right] - i\mu \int_0^t dt' \zeta(t') \dot{Q}'(t')\, , \nonumber \\
  \label{inffct}
\end{eqnarray}
where the last term  accounts for the initial preparation of the bath\cite{lucke}.
 Here, one sums over all paths which connect $\eta(0), \zeta(0)$ with $\eta(t), \zeta(t)$ where the initial values have to be chosen according
  to the initial preparation (\ref{eq:initialprep}) and the final endpoints according to the measured observable (\ref{observables}).
 The kernel $Q=Q'+ i Q''$ is related to the bath correlation $\ddot{Q}(t)=\langle{\cal E}(t){\cal E}(0)\rangle/\hbar^2$ and reads
\begin{equation}\label{qt}
Q(t) = \int_0^{\infty} d\omega \frac{J(\omega)}{\pi\omega^2} \Big[\coth\left(\frac{\hbar\beta\omega}{2} \right)
         (1-\cos\omega t) + i\sin\omega t \Big].
\end{equation}

The most efficient methodology to evaluate the above path integrals non-perturbatively  is  the path integral Monte Carlo approach (PIMC) \cite{egger,muehlbacher1}. The procedure starts from a Trotter-Suzuki discretization of the total time interval $t$ with time increment
$\tau=t/q$ and Trotter number $q$. The full expression (\ref{pathinte})   reads with $P_\nu^i \equiv P_\nu((i-1)\tau)$
\begin{eqnarray}
P_\nu^i
       &=& \sum_{\substack{\zeta_{1...q+1}\\ \eta_{1...q+1}}} \rho_{0}(\eta_1,\zeta_1)
           \left(\prod_{j=1}^{q} KK(\zeta_j, \zeta_{j+1})_{\eta_j, \eta_{j+1}} \right) \nonumber \\
       &&  \times M_{\nu, i}[\zeta,\eta] \exp(-\Phi[\zeta, \eta])\, .
       \label{MCobs}
\end{eqnarray}
 with matrices $KK_{\eta_i, \eta_{i+1}}$ containing the free propagation of the TLS and $M_{\nu, i}$ being the measurement operator for the observable $\sigma_\nu$ at the time step $(i-1)\tau$ (for details see\cite{muehlbacher1} and  the Appendix).

The impact of the reservoir is completely determined by the spectral density. A common form is given by
\begin{eqnarray}
   J_s(\omega) = 2\pi\alpha \omega_c^{1-s}\omega^s \exp(-\omega/\omega_c)
   \label{specdens}
\end{eqnarray}
where $\alpha$ is a dimensionless coupling strength, $\omega_c$ a soft ultra-vilolett-cutoff, and $s$ the spectral exponent. For $s=1$, the
bath is called Ohmic, while for $0<s<1$ it is referred to as sub-Ohmic. In both cases, analytical expressions for $Q(t)$ are known.

The numerical evaluation of the expression (\ref{MCobs}) by means of Monte Carlo methods is hampered by the so-called dynamical sign problem which originates from the interference structure of quantum mechanics and leads to an exponential decrease of the signal to noise ratio for longer times. In the past, several techniques have been developed to substantially soothe its impact on PIMC simulation  \cite{egger,muehlbacher1}. The general idea is to reduce the sampling space by integrating out exactly large parts of the configuration space, i.e., to sum over the paths $\eta$ in  (\ref{MCobs}). Only the remaining $q$-fold sum over the relative coordinates $\zeta_{1\leq i\leq q}$ is performed by a Metropolis sampling.
This way, simulations yield numerically exact results up to times where equilibration sets in with error bars of typically less than 1\%. We will discuss below a procedure how to further increase the simulation time by linking segments of PIMC on moderate time scales.

\section{Analytical Treatment}\label{niba}

An approximate and well-studied treatment of the SBM which captures also strong coupling and low temperatures is
the noninteracting blip approximation (NIBA) \cite{weiss}. It is based on a separation of scales between times where $\rho(t)$ resides in a diagonal state (sojourn) and those where it is in an off-diagonal state (blip). In the NIBA only reservoir induced intra-blip and blip-proceeding sojourn interactions in the influence functional (\ref{inffct}) are kept, while remaining correlations are dropped.
 Although this approximation is expected to fail in regions of parameter space with
 long memory effects, it provides in many cases quantitatively accurate results and in others at least a qualitative  description of the dynamics. For further details we refer to the extensive literature \cite{weiss,grifoni1,grifoni2,grifoni3}.

For the time dependent polarization $P_z(t)=\langle \sigma_z(t)\rangle$  with the initial preparation $P_z(0)=\mu=1$ one derives
\begin{eqnarray}\label{niba-pz}
  \frac{dP_z(t)}{dt} = - \int_0^t du \; K(t-u) P_z(u)
\end{eqnarray}
with kernel
\begin{eqnarray}
   K(t) & = & \Delta^2 e^{-Q'(t)} \cos[Q''(t)]\, .
   \label{NIBA_z}
\end{eqnarray}
In ranges of parameters space where $K(t)$ is sufficiently short ranged, the dynamics (\ref{niba-pz}) can be approximated by
\begin{equation}\label{pzinco}
\frac{d P_z(t)}{dt}\approx -\Gamma_{\rm NIBA}\, P_z(t)
\end{equation}
with a rate $\Gamma_{\rm NIBA}=\int_0^\infty du \; K(u)$ describing monotone decay.

Further, for the imaginary part of the coherences one has $P_y(t)=-\dot{P}_z(t)/\Delta$, while the real part $P_x$ follows for a symmetric system from
\begin{equation}
  P_x(t) = \Delta {\displaystyle \int_0^t du \; e^{-Q'(u)} \sin Q''(u)}\, .
  \label{NIBA_x}
\end{equation}

\section{Ohmic Spectral Density}\label{ohmic}

As mentioned above, a major goal of this work is to explore the options arising when there is also access to off-diagonal elements of the reduced density operator and when using generalized initial states (\ref{eq:initialprep}).  We start with some results for an Ohmic environment as a test-bed for the more challenging sub-Ohmic case considered afterwards.  First, PIMC data for $P_x$ are presented and compared to NIBA predictions. This provides the basis for a new technique to extend the simulation time for PIMC.

\subsection{Off-diagonal elements}

It is well-known that the Ohmic SBM displays damped oscillatory dynamics in $P_z(t)$  for weaker couplings and/or lower temperatures (coherent domain) and incoherent decay for stronger friction and/or higher temperatures\cite{leggett,weiss}. At zero temperature, this transition occurs at $\alpha=1/2$ for $\omega_c\gg \Delta$. These and other properties have been studied numerically by several approaches including PIMC and also in comparison with NIBA. Less attention has been paid to $P_x$ though, for which it is believed that NIBA predictions are less reliable. For example,  in the incoherent regime, Eq. (\ref{NIBA_x}) predicts at $T=0$ a finite value
\begin{equation}
  P_x (t\to \infty) = \frac{\Delta}{\omega_c} \frac{1}{2\alpha-1}
\end{equation}
for $\alpha>1/2$ while it diverges for $\alpha \leq 1/2$.

The implementation of $\sigma_x$ as measurement operator in the PIMC is not trivial. Namely, the approach is only efficient when one extracts observables at intermediate times $u$ from a {\em single run} over the {\em total time } $t\geq u$. Hence, since the influence functional is non-local in time, the effective measurement operator in the PIMC corresponding to $\sigma_x$ must also include contributions from the bath\cite{kast} (see also Appendix).

\begin{figure}
\epsfig{file=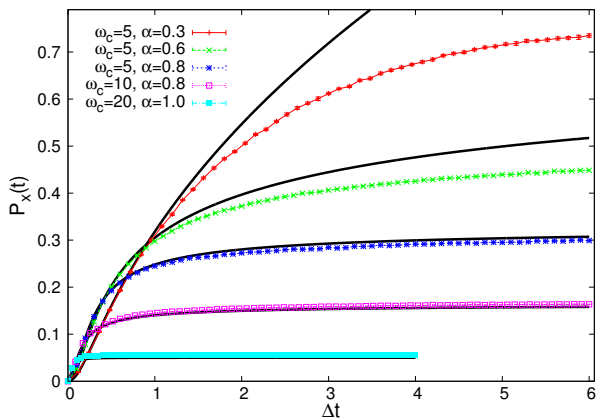, angle=270, width=7.5cm}
\vspace*{0.5cm}
\caption{Dynamics of $P_x(t)$ for  Ohmic spectral density at $T=0$. PIMC results (dots with error bars)
and analytical NIBA results (black lines) for various values of the coupling $\alpha$ and the cut-off $\omega_c$.}
\label{sigma_x-Ohm}
\end{figure}

 Fig. \ref{sigma_x-Ohm} shows a comparison between PIMC data and the NIBA at zero temperature.
A polarized initial preparation is chosen with $P_z(0)=1$. For longer times, $P_x(t)$ indeed increases stronger for smaller values of the coupling, while it quickly saturates for stronger damping. In this latter range ($\alpha\gtrsim 0.8$), NIBA is in excellent agreement with the data. For $0.8\gtrsim\alpha\gtrsim0.5$, i.e., in the incoherent domain, it still provides qualitatively correct results over the time period considered, but fails completely in the coherent range $\alpha<0.5$. Further, the larger the cut-off, the better is the agreement with the PIMC simulations. Increasing $\alpha$ and $\omega_c$ tends to suppress coherences and thus pushes the system towards a localized state which exists for $\alpha>1$.

For finite temperatures $T>0$, the dynamics of $P_x(t)$ at moderate damping remains  qualitatively the same (not shown) and is strongly suppressed at high temperatures when $Q'$ dominates against $Q''$ in (\ref{NIBA_x}).

\subsection{Long-time simulations: CH-PIMC}

The time dependent density operator of the full compound obeys the semi-group property
\begin{eqnarray}
W(t)&=& \, U(t,t_s)\, W(t_s)\, U^\dagger(t,t_s)\nonumber\\
&=& U(t,0)\, W(0)\, U^\dagger(t,0)\,  \nonumber\\
\end{eqnarray}
with $U(t,t')=\exp[-i H (t-t')/\hbar)$.
In general, this is not true for the reduced density operator (\ref{reduced}). However, there are two domains in parameter space where this symmetry holds at least approximately. One is the domain of very weak coupling, sufficiently high temperatures, and large cut-off frequencies, i.e., $\alpha \Delta\hbar\beta\ll 1$ and $\Delta\ll \omega_c$. Then, a perturbative treatment including a coarse graining in time (Born-Markov approximation) leads to master equations for $\rho(t)$ which are local in time and are determined by a time-independent generator. In the other domain, coherences (off-diagonal elements) in $\rho(t)$ are suppressed due to the bath such that effectively the full density operator at intermediate times takes the form $W(t_s)\approx \rho(t_s)\, \exp[-\beta (H_B-\mu_s\, {\cal E})]/Z_B$. This can be interpreted as a quasi-classical regime, where, however, temperature does not need to be high, but rather friction sufficiently strong. Accordingly, quantum effects may still substantially influence the relaxation dynamics towards thermal equilibrium. While corresponding rates are of great interest experimentally, to extract them from PIMC data, typically requires very expensive long times simulations.

Here we focus on this latter regime, not accessible by master equations, and develop a PIMC scheme which exploits the fact that a semi-group property applies approximately.
The proposed scheme works for $P_z(t)$ as observable as follows. One defines auxiliary densities $\rho_n(t)={\rm Tr}_B\{W_n(t)\}$ on segments $t\in [n t_s, (n+1) t_s]\, , n=0, 1, 2, \ldots N$, where
\begin{equation}
W_n(t)=U(t,n t_s)\ \rho_n(n t_s) \frac{1}{Z_B} {\rm e}^{-\beta (H_B-\mu_n {\cal E})}\ U^\dagger(t,n t_s)\, .
\end{equation}
and $t_s$ is sufficiently large compared to the bath memory time.
In
\begin{equation}
\rho_n(n t_s)= \frac{1}{2} \left(\mathds{1} + \sum_{\nu=x,y,z} P_{\nu, n}(n t_s)\,  \sigma_\nu \right)\ , n\geq 1,
\end{equation}
the values for $P_{\nu,n}(n t_s)$ as well as the bath-shift $\mu_n$ are adjusted such that one achieves a smooth matching according to
\begin{eqnarray}
P_{z, n}(n t_s) &=& P_{z, n-1}(n t_s)\ , \ \dot{P}_{z, n}(n t_s) = \dot{P}_{z, n-1}(n t_s)\ , \nonumber\\
\ddot{P}_{z, n}(n t_s) & = &\ddot{P}_{z, n-1}(n t_s)\, .
\end{eqnarray}
For $n=0$, one has the initial density $\rho_0(0)=|+1\rangle\langle+1|$ with $\mu_0=1$.
 Now, due to continuity and the exact relation $P_y(t)=- \dot{P}_z(t)/\Delta$, the first two conditions fix the elements $P_{z,n}(n t_s)$ and $P_{y,n}(n t_s)$ through the time evolution over the previous segment. To satisfy the third condition one properly adjusts $P_{x, n}(n t_s)=\delta P_{x,n}+P_{x, n-1}(n t_s)$ and the bath shift $\mu_n$. The reservoir distribution is thus progressively distorted by the spin dynamics and the reduced density carries information about how the bath modes induce the spin relaxation.
 This way, the correct dynamics over the total time interval $t_{\rm tot}=(N+1) t_s$
is approximated by a chain of PIMC simulations over time intervals $t_s$ through
\begin{equation}
P_z(t)\simeq P_{z, n}(t)\ \ { \rm for}\ t\in [n t_s, (n+1) t_s], \ n=0, 1,\ldots N\, .
\end{equation}
Typical lengths of the segments range from a few up to 15 times $1/\Delta$.
The validity of this Chain-PIMC (CH-PIMC) is well-controlled. If for the combined data one obtains a smooth matching it works, if ''kinks'' appear, it does not.  In the moderate to strong friction regime, this implies $P_x\ll 1$ which can be verified at each intermediate time step (for details see below).
Note that the procedure is exact at $\alpha=0$ and thus expected to cover also the domain of very weak coupling.
 The computational gain is substantial: As the dynamical sign problem increases exponentially with $t$, the CPU time for a simulation over the total time interval $t=n\, t_s$ scales as ${\rm e}^{a n t_s}$ ($a>0$ is a constant) while for a chain with  $n$ segments of length $t_s$ one has an algebraic increase $n\, {\rm e}^{a t_s}$.

\begin{figure}
\epsfig{file=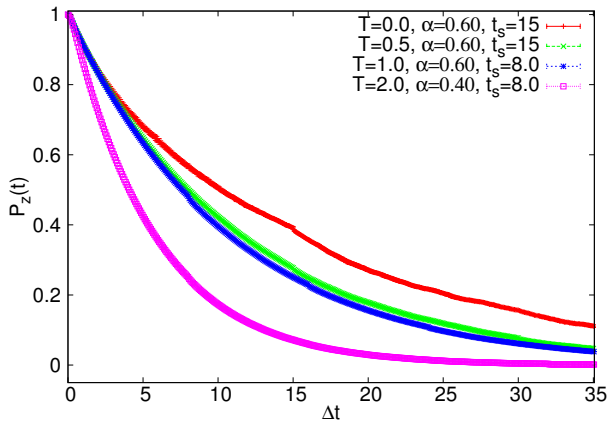, angle=270, width=7.5cm}
\vspace*{0.5cm}
\caption{Dynamics of $P_z(t)$ according to the CH-PIMC with optimized matching for various coupling strengths and temperatures. For each parameter set a different time interval $t_s$ was used. See text for details. Temperatures are scaled in units of $\hbar\Delta/k_B$.}
\label{cutting}
\end{figure}

\begin{figure}
\epsfig{file=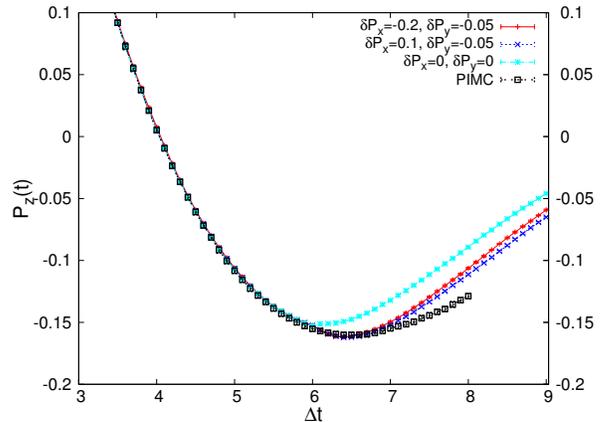, angle=270, width=7.5cm}
\vspace*{0.5cm}
\caption{Dynamics for $P_z(t)$ in the coherent regime ($\alpha=0.3$) using the CH-PIMC for various deviations $\delta P_x$, $\delta P_y$. Parameters are $T=0$ and $t_s=6/\Delta$.}
\label{dxdy_coh}
\end{figure}

In Fig.~\ref{cutting} results for the CH-PIMC are shown for Ohmic friction and larger couplings and over a wide temperature range. Here, we found within the statistical errors an optimal matching for $\mu_n=P_{z, n}(n t_s)$ and $\delta P_{x, n}\approx 0$. Hence, at each intermediate time step the drag of the system onto the equilibrium position of the bath is updated which accounts for dynamically induced system-bath correlations. The procedure also works in the domain of very low temperatures where the relaxation dynamics is strongly influenced by quantum effects but only as long as $P_x(n t_s)\lesssim 0.1 $. At zero temperature off-diagonal elements tend to be considerably larger, namely, $P_x(n t_s)\approx 0.3 $, and ''kinks'' at the intermediate time steps cannot be removed by varying $\delta P_{x, n}$ and $\mu_n$.
In the domain $\Delta\hbar\beta \lesssim 2$ and $\alpha>0.5$  (lower temperatures, moderate to strong damping) one has thus access to the full equilibration dynamics where approximate treatments fail such as e.g. weak coupling master equations.
When the reduced dynamics is determined by strong coherences, but friction is not extremely weak, the CH-PIMC fails as seen in Fig.~\ref{dxdy_coh}. We recall that the CH-PIMC describes the bare coherent dynamics ($\alpha=0$) exactly though.

\section{Sub-Ohmic heat bath}\label{subohmic}

The methodology developed in the previous section is now used to analyze the SBM with sub-Ohmic spectral densities, i.e., densities (\ref{specdens}) with spectral exponents $0<s<1$. This class of reservoirs has received considerable attention recently as it seems to constitute the dominant noise source in solid state devices at low temperatures such as superconducting qubits \cite{fnoise} and quantum dots \cite{ring} with the spectral exponent $s$ determined by
the microscopic nature of environmental degrees of freedom. It also appears in the context of quantum impurity systems \cite{heavyf} and nanomechanical oscillators \cite{nanomech}.

\begin{figure}
\epsfig{file=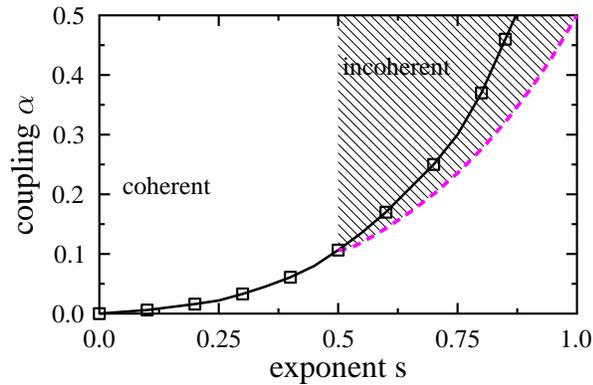, angle=0, width=8cm}
\caption{Domains of coherent (white) and incoherent (shaded) dynamics of the SB model for a sub-Ohmic environment (\ref{specdens}) at zero temperature. Above (below) the solid line $\alpha_c(s)$ the system asymptotically reaches a thermal equilibrium which is localized (delocalized) \cite{subohmic2}. A coherent-incoherent changeover only occurs along the dashed line $\alpha_{CI}(s)$ \cite{kast}.}\label{phasediagram}
\end{figure}

The model is further of fundamental interest with respect to its thermodynamical as well as its non-equilibrium properties. In thermal equilibrium at zero temperature a dissipation induced quantum phase transition  occurs \cite{subohmic1,subohmic2,subohmic3,subohmic4}. Accordingly, at critical coupling strengths $\alpha_c(s)$ the system exhibits a transition from a delocalized state [$P_z(t\to \infty)=0$] with tunneling between the two spin orientations (weak friction) to a localized one [$P_z(t\to \infty)\neq 0$] with basically classical behavior (stronger friction) occurs. With respect to the relaxation dynamics,
it has been shown recently \cite{kast} that there is a coherent-incoherent transition in the regime $1/2\leq s<1$ at coupling strengths $\alpha_{CI}(s)\leq \alpha_c(s)$, while such a transition is absent with coherences surviving even for ultra-strong coupling
 in the domain $0<s<1/2$ (cf.~Fig.~\ref{phasediagram}). In this regime, damped oscillations occur in $P_z(t)$ with frequencies that are in agreement with those extracted from the NIBA
\begin{equation}
   \Omega_{s} \approx \Lambda_{s} \approx \frac{2\alpha \omega_c}{s}\, .
   \label{omegaeff}
\end{equation}
Here $\Lambda_s=\int_0^\infty d\omega J_s(\omega)/(\pi \omega)= 2\alpha\omega_c |\Gamma(s-1)|$ is the reorganization energy.
Likewise, the damping rate is given by $\gamma_s(\alpha)\approx s\, \Lambda_s$ such that $\Omega_s/\gamma_s\gg 1$ (underdamped regime).
We note in passing that for a pure dephasing model ($H_{TLS}=\frac{\Delta}{2}\sigma_z$) which can be solved
analytically coherent oscillations persist up to arbitrary strong couplings below a critical $s<s_c$ \cite{nalbach2013}.

In this section we will explore first to what extent coherences are robust against finite temperatures and various cut-off procedures to better understand the domain in parameter space where the CH-PIMC is expected to work. Second, in the so determined incoherent regime relaxation rates are obtained from CH-PIMC data and compared to NIBA predictions.

\subsection{Coherent regime}

The central ingredient in the influence functional (\ref{inffct}) which captures the impact of the reservoir is the bath correlation (\ref{qt}).
For sub-Ohmic spectral densities of the form (\ref{specdens}) it can be calculated analytically to read $Q=Q_0+Q_\beta$ with the zero temperature part $Q_0(t)=2\alpha \Gamma(s-1) [1-z^{1-s}]$, where $z=1+i\omega_c t$, and a part for finite temperatures
\begin{eqnarray}\label{corrT}
Q_\beta(t)&=&2\alpha \Gamma(s-1) \kappa^{1-s} [2\zeta(s-1, 1+1/\kappa)\nonumber\\
&-&\zeta(s-1,1+z/\kappa)-\zeta(s-1,1+z^*/\kappa) ]
\end{eqnarray}
with the Hurwitz function $\zeta(z,a)$ and $\kappa=\hbar\beta\omega_c$.

Now, according to the NIBA expression (\ref{niba-pz}) with $Q'=Q_0'+Q_\beta$ and $Q''=Q_0''$, finite temperatures lead to an exponential suppression of the integral kernel and have thus the tendency to destroy coherences (oscillations) in the relaxation dynamics of $P_z(t)$. In particular, in the domain $s\ll 1$ and for temperatures $\omega_c\hbar\beta\gg 1$, one has $Q'(t)\approx Q_0'(t) [1+t/(s\, \hbar\beta)]$ so that on the time scale of the bare dynamics $1/\Delta$ thermal fluctuations are relevant if $\Delta\hbar\beta \lesssim 1/s$. Then, due to the accumulation of low frequency bath modes, the SBM becomes progressively sensitive to thermal fluctuations for smaller $s$.

\begin{figure}
\epsfig{file=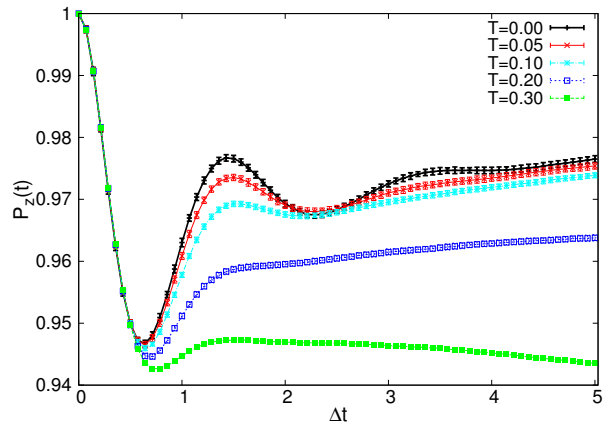, angle=270, width=7.5cm}
\vspace*{0.5cm}
\caption{Dynamics of $P_z$ for various temperatures for $s=1/4$ and $\alpha=0.1$. The initial bath preparation is $P_z(0)=1$.}
\label{finite_T}
\end{figure}

This is illustrated in Fig. \ref{finite_T} which displays the changeover from oscillatory to overdamped motion at $s=1/4$ with increasing temperature. Indeed, for temperatures $k_{\rm B} T\approx s\, \hbar\Delta$ coherences are washed out and after a transient period of time the dynamics turns into a monotone relaxation described by a single time constant, the relaxation rate. It is this rate that we extract below within the CH-PIMC.

\begin{figure}
\epsfig{file=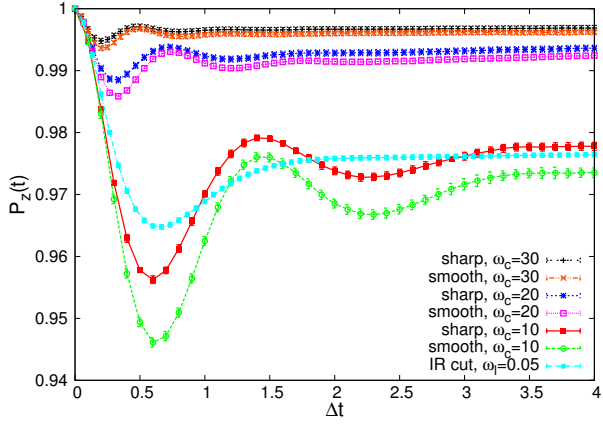, angle=270, width=7.5cm}
\vspace*{0.5cm}
\caption{Dynamics of $P_z(t)$ for various cut-off schemes: a sharp cut-off (red, green, blue) and a soft cut-off (light blue, yellow, black) for the ultra-violet cut-off are shown together with the situation of an additional infrared cut-off (magenta); see text for details.}
\label{cutoff}
\end{figure}

Before we do so, however, let us analyze the sensitivity of the coherent dynamics on the cut-off procedure. According to (\ref{omegaeff})
it is mainly determined by the reorganization energy $\Lambda_s$. The corresponding integral is dominated by the low frequency profile of the spectral density $J_s(\omega)$. As a consequence, for sufficiently large cut-off frequencies $\omega_c/\Delta\gg 1$ dynamical properties of $P_z(t)$ are basically independent of the specific form of the cut-off, e.g., a smooth exponential as in (\ref{specdens}) or a sharp cut-off as in $\tilde{J}_s(\omega)= 2\pi\alpha \omega_c^{1-s}\omega^s \, \Theta(\omega_c-\omega)$. This is in contrast to an infrared cut-off $\omega_l$ which has pronounced effect on coherences as seen in Fig.~\ref{cutoff} for $T=0$. Even a small $\omega_l/\Delta=0.05$ pushes the dynamics toward an incoherent decay similar to that at elevated temperatures, while varying the ultra-violet cut-off scheme has almost no influence but only leads to a renormalization of the tunnel coupling \cite{weiss}.
To summarize this analysis, we find incoherent relaxation dynamics in the sub-Ohmic SBM for finite temperatures also in the domain $0<s<1/2$. The same is true for a low frequency cut-off, while the dynamics is robust against various high cut-off schemes.

\subsection{Incoherent regime}

\begin{figure}
\epsfig{file=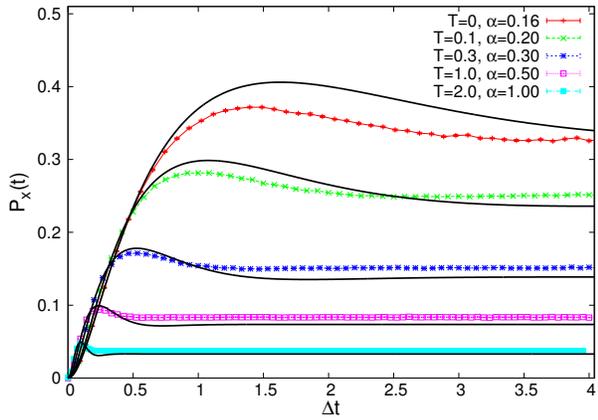, angle=270, width=7.5cm}
\vspace*{0.5cm}
\caption{Dynamics of the coherences $P_x(t)$ in a sub-Ohmic bath with $s=1/2$ for various temperatures and coupling strengths.
Symbols refer to PIMC data whereas solid lines show the corresponding NIBA results according to (\ref{NIBA_x}).}
\label{sigmax-SO}
\end{figure}

In this regime, we start as above for the Ohmic SBM with the dynamics of $P_x(t)$ in Fig.~\ref{sigmax-SO} for various temperatures and coupling strengths. As expected, stronger friction and finite temperatures lead to a suppression of off-diagonal elements of the reduced density and thus drive the system towards a classical distribution. The NIBA expression (\ref{NIBA_x}) captures the PIMC data qualitatively correctly, but quantitative agreement is only found for sufficiently high temperatures and stronger couplings. In the low temperatures range, the NIBA predictions provide a reliable estimate only for long times. Further, according to Fig.~\ref{phasediagram} and the data in Fig.~\ref{sigmax-SO},
one expects the CH-PIMC to be applicable for $s\geq 0.5$ and sufficiently high temperature, i.e.\ $\Delta \hbar \beta \lesssim 2$.

\begin{figure}
\epsfig{file=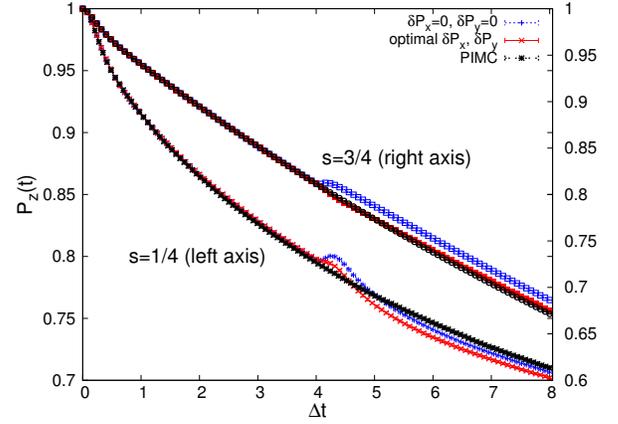, angle=270, width=7.5cm}
\vspace*{0.5cm}
\caption{Polarization $P_z(t)$ for sub-Ohmic spectral density at $T=\hbar\Delta/k_B$ with $s=3/4$, $\alpha=0.5$ and $s=1/4$, $\alpha=0.1$  (see also Fig. \ref{phasediagram}) using the CH-PIMC for various deviations $\delta P_x$, $\delta P_y$.}
\label{piecingon-S0}
\end{figure}

This is illustrated in Fig.~\ref{piecingon-S0}, where the relaxation dynamics for the polarization $P_z(t)$ is shown. Due to the finite temperature, $P_z(t)$ decays in both cases monotonically but in the regime of small $s$-values, where at $T=0$ coherences survive up to arbitrary strong coupling (cf.~Fig.~\ref{phasediagram}), off-diagonal elements are still substantial at weak friction. One also finds that for a sub-Ohmic bath optimal matching is achieved for $\mu_n=1$, i.e. the initial bath preparation. The reason for that is the sluggish bath with a large portion of low frequency bath modes that remain basically static on the time scale of a few $1/\Delta$.

\begin{figure}
\epsfig{file=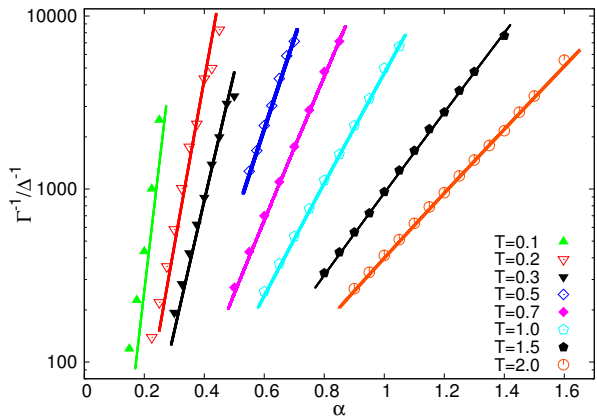, angle=270, width=7.5cm}
\vspace*{0.5cm}
\caption{Inverse equilibration rates (lifetimes) in the incoherent regime for $s=1/2$ as a function of
$\alpha$ and $T$ (in units of $\hbar\Delta/k_{\rm B}$). NIBA predictions (lines) according to (\ref{NIBArates}) are shown together with data extracted from standard PIMC ($T\leq 0.3$) and CH-PIMC ($T\geq 0.5$) simulations.}
\label{rates}
\end{figure}

In the range of spectral densities where the CH-PIMC applies, it can now be used to extract from the long time dynamics relaxation rates. 
However, as the required time disctretization $q/t$ is much smaller for low temperatures and weak couplings, standard PIMC results are already sufficient to fit the rates in the regime where CH-PIMC usually fails, i.e. $\Delta\hbar\beta \gtrsim 2$. 
Our results can also be used to test the accuracy of the NIBA [cf.~(\ref{pzinco})]. In fact, an analytical evaluation \cite{leggett} yields the compact expression $\Gamma_{\rm NIBA}=a_s \exp\left[ -2\alpha b_s \Gamma(s-1)\right] $ with
\begin{eqnarray}
   b_s & = & 1+2(\omega_c \hbar \beta)^{1-s} [\zeta(s-1,1)-\zeta(s-1,1/2)] \nonumber \\
   a_s & = & \frac{\Delta^2}{2\omega_c}
             \left[\frac{\pi (\omega_c \hbar \beta)^{1+s}}{2\alpha \Gamma(1+s) \zeta(1+s, 1/2)} \right]^{1/2}\, .
   \label{NIBArates}
\end{eqnarray}
Apart from the rather weak $\alpha$-dependence in the pre-factor, the decay rate depends exponentially on the coupling strength.
This is depicted in Fig. \ref{rates} with an almost linear behavior of $\log(1/\Gamma)$ as a function of $\alpha$ for different temperatures. The NIBA gives indeed a perfect description over a broad range of parameters. The overall magnitude of the lifetimes and thus the time scale for equilibration is extremely large compared to the bare dynamics (see also \cite{subohmic3}). Effectively, the spin is then almost frozen and displays a slow equilibration which may be beyond typical experimental time scales. These results reveal the benefit of the CH-PIMC, namely, to capture the domain of long times and strong coupling. One may argue that we test the validity of an approximate analytical treatment (NIBA) by comparing its predications with those from an approximate numerical treatment (CH-PIMC). To avoid any inconsistencies we thus also performed full PIMC simulations and compared them with CH-PIMC results (not shown). The latter are again in perfect agreement within the statistical error with exact data in all cases where the matching procedure yields smooth relaxation dynamics.

\subsection{Entropy and phase transition}

With the full reduced density operator at hand, an interesting quantity to measure with PIMC simulations is the von Neumann entropy
\begin{eqnarray}
   S/k_{\rm B} = -{\rm Tr} (\rho \log \rho) = -p_+ \log p_+ - p_- \log p_-
   \label{neumann}
\end{eqnarray}
where $p_{\pm}=\frac{1}{2}\left(1\pm\sqrt{\langle \sigma_x \rangle^2 + \langle \sigma_y \rangle^2 + \langle \sigma_z \rangle^2}\right)$.
It is known to provide at zero temperature the entanglement between system and reservoir for the SBM\cite{hofstetter}. In particular,
in case of the sub-Ohmic SBM it was shown previously that at $T=0$ the quantum phase transition in thermal equilibrium from the delocalized to the localized phase corresponds to a maximum of the entropy \cite{hofstetter,plenio}. Here we use PIMC techniques to explore to what extent
  dynamical information on finite time scales allows to detect this transition.

\begin{figure}
\epsfig{file=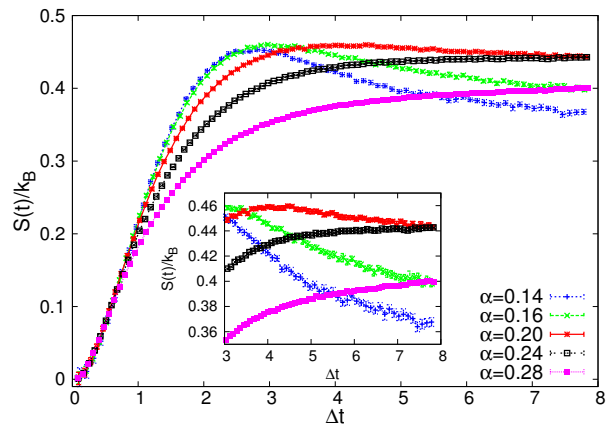, angle=270, width=7.5cm}
\vspace*{0.5cm}
\caption{Entropy for various couplings $\alpha$ at $s=0.75$ and $T=0$. The inset shows a blow-up for longer times. See text for details.}
\label{PT-s075}
\end{figure}

We start in Fig.~\ref{PT-s075} with a spectral density with exponent $s=3/4$. Apparently, towards the end of the simulation time range the entropy tends to saturate for increasing coupling strength. The PIMC simulations are not fully equilibrated, but from the data one can at least estimate a maximum in the entropy to occur for $0.3\gtrsim\alpha>0.24$ with the correct value \cite{subohmic2} being $\alpha_c(s=0.75)=0.30$. For longer simulation times the dynamical sign problem drastically deteriorates the data.

\begin{figure}[htb]
\epsfig{file=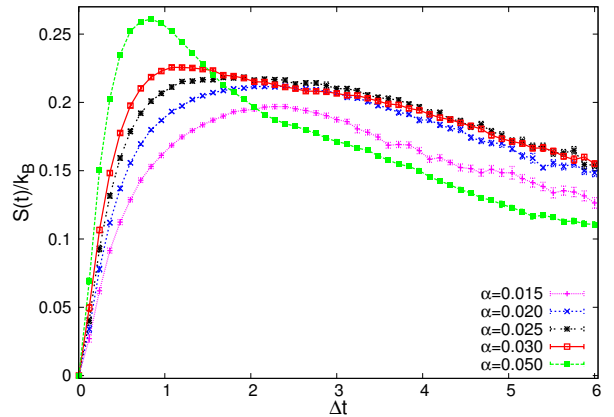, angle=270, width=7.5cm}
\vspace*{0.5cm}
\caption{Same as in Fig. \ref{PT-s075} but for $s=0.25$ with an initial preparation $P_x(0)=0.8$, $\mu=P_z(0)=0.6$, $P_y(0)=0$.}
\label{PT-s025}
\end{figure}

This problem becomes more severe for smaller exponents $s\leq 1/2$ where PIMC simulations become increasingly  demanding due to  longer saturation time scales.
In order to reduce the transient regime, we start not with a polarized initial state but from a density $\rho_0(0)$ with elements $P_\nu$ chosen closer to known thermal equilibrium values\cite{hofstetter}, i.e., $\mu=P_z(0)=0.6$, $P_x(0)=0.8$ and $P_y(0)=0$. As seen for $s=0.25$ in  Fig. \ref{PT-s025}, this procedure leads on shorter time scales to a maximum of the entropy within the range $0.02<\alpha<0.025$. This is in full agreement with the exact value $\alpha_c(s=0.25)\approx 0.022$.

\section{Conclusions}\label{conclusions}

In this work the time evolution of the reduced density of a spin-boson model has been investigated within numerically exact PIMC simulations in domains of strong coupling to environmental degrees of freedom and also at elevated temperatures. An efficient formulation has been developed to extract simultaneously information about populations and coherences within a single PIMC run. With the full density matrix at hand, segments of PIMC simulations on moderate time scales could be combined based on successively determined initial states. This method (CH-PIMC) allows to cover long times and to extract e.g.\ equilibration rates which are not accessible by perturbative treatments. Specific applications for spin-boson models with sub-Ohmic spectral densities reveal the domains in parameter space where the method is reliable. Where it does not work, it gives insight into the strength of long-ranged system-bath correlations.

\acknowledgments
We thank J. Stockburger for fruitful discussions. Financial support was provided by the German-Israeli Foundation, the
DFG through SFB/TRR21 and AN 336/7-1.

\appendix*

\section{Implementation of the operator for coherences}\label{appendix}

The PIMC is only efficient when one extracts observables at intermediate times $0\leq u$ from a {\em single} run over a total time $t\geq u$. Since the influence functional is non-local in time, the measurement operator in the PIMC corresponding to $\sigma_x$ includes also contributions from the bath. One starts with a Trotter-Suzuki discretization of the time axis with $q$ Trotter steps according to $t=q\tau$ with time increment $\tau$. The discretized form of the influence functional (\ref{inffct}) in the original forward and backward paths $\sigma$ and $\sigma'$ reads \cite{muehlbacher1}
\begin{eqnarray}
   \Phi & = & \sum_{j=2}^q  \zeta_j\sum_{k=2}^{j} \Lambda_{j-k} \zeta_k + \zeta_1 \sum_{j=2}^q \Lambda_j^{(1)} \zeta_j
            + \sum_{j=2}^q  \zeta_j\sum_{k=2}^{j-1} X_{j-k} \eta_k \nonumber \\
         && + i \eta_1 \sum_{j=2}^q X_j^{(1)} \zeta_j + i \sum_{j=1}^q X_j^{(\mu)} \zeta_j
         \label{inffct_disc}
\end{eqnarray}
with
\begin{eqnarray}
\Lambda_0 & = & {\rm Re}\;[Q(\tau)], \nonumber \\
\begin{array}{c} \Lambda_{1\leq j\leq q-2} \\ X_{1\leq j\leq q-2} \end{array} & = &
              \begin{array}{c} {\rm Re} \\ {\rm Im} \end{array}
              \begin{array}{l} [ Q((j+1)\tau)-2Q(j\tau) \\  \hspace{13mm} + Q((j-1)\tau) ] \end{array} \nonumber \\
\Lambda_1^{(1)} & = & {\rm Re}\;[Q(\tau/2)], \nonumber \\
\begin{array}{c} \Lambda_{2\leq j\leq q}^{(1)} \\ X_{2\leq j\leq q}^{(1)} \end{array} & = &
              \begin{array}{c} {\rm Re} \\ {\rm Im}\end{array}
              \begin{array}{l} [Q((j-1/2)\tau)-Q((j-1)\tau) \\
                               \;\; -Q((j-3/2)\tau)+Q((j-2)\tau)],
              \end{array} \nonumber \\
X_1^{(\mu)} & = & -\mu {\rm Im}\;[Q(\tau/2)], \nonumber \\
X_{2\leq j\leq q}^{(\mu)} & = & -\mu {\rm Im}\;[Q((j-1/2)\tau) - Q((j-3/2)\tau)]. \nonumber \\
\end{eqnarray}
Since paths are closed at the end of the full simulation period $t$ one has $\zeta_{q+1}=0$.
Further, according to the chosen path discretization, the spin flips occur at $t_j = \left(j-\frac{1}{2}\right)\tau$.

Now, let us first explain the procedure for vanishing coupling to the bath $\alpha=0$. Populations $P_z^i \equiv P_z((i-1)\tau)$
are then simply obtained from
\begin{eqnarray}
   P_z^i & = & \sum_{\substack{\sigma_1,...,\sigma_i \\ \sigma'_1,...,\sigma'_i}}
                    \left[\prod_{j=1}^{i-1} K^*(\sigma'_j, \sigma'_{j+1}) K(\sigma_{j+1}, \sigma_j) \right] \times \nonumber \\
                 && \times \rho_0(\sigma_1, \sigma'_1) \delta_{\sigma_i, \sigma'_i} (\delta_{\sigma_i, 1} - \delta_{\sigma_i, -1})\, ,
            \label{sigzinter}
\end{eqnarray}
where matrices $K(\cdot,\cdot)$ represent the short time propagators of the TLS and $*$ means complex conjugation.
This expression can be extended to an expression for paths over the full time interval $0\leq u\leq t$ by inserting
into (\ref{sigzinter}) the identity
\begin{eqnarray}
   \delta_{\sigma_i,\sigma'_i} & = & \sum_{\substack{\sigma_{i+1},...,\sigma_{q+1} \\ \sigma'_{i+1},...,\sigma'_{q+1}}}
           \left[\prod_{j={i}}^{q} K^*(\sigma'_j, \sigma'_{j+1}) K(\sigma_{j+1}, \sigma_j) \right] \nonumber \\
     &&    \times \delta_{\sigma_{q+1}, \sigma'_{q+1}}\, .
\end{eqnarray}
Accordingly, $P_z^{i}$ at each intermediate time step can be read out from a Monte Carlo run over the whole time interval
$0\leq t'\leq t$ by applying the operator $M_{z,i}^{\alpha=0}= \delta_{\sigma_i, \sigma'_i} (\delta_{\sigma_i, 1} -
\delta_{\sigma_i, -1})$.

For off-diagonal elements $P_x^i \equiv P_x((i-1)\tau)$ things become a bit more complicated.
To extend the expression for intermediate times
\begin{eqnarray}
   P_x^i & = & \sum_{\substack{\sigma_1,...,\sigma_i \\ \sigma'_1,...,\sigma'_i}}
         \left[\prod_{j=1}^{i-1} K^*(\sigma'_j, \sigma'_{j+1}) K(\sigma_{j+1}, \sigma_j) \right] \nonumber \\
         && \times \rho_0(\sigma_1, \sigma'_1) \delta_{\sigma_i, \sigma'_i} (\delta_{\sigma_i, 1} - \delta_{\sigma_i, -1})
   \label{sigma_x1}
\end{eqnarray}
such that one can extract expectation values of $\sigma_z$ {\em and} $\sigma_x$ from a single PIMC run, one has to propagate
artificially to a diagonal state at $i+1$.
Thus, one multiplies (\ref{sigma_x1}) with
\begin{equation}
   1 = \frac{1}{2} \sum_{\sigma_{i+1}, \sigma'_{i+1}} \delta_{\sigma_{i+1}, \sigma'_{i+1}}
       \frac{K^*(\sigma'_{i}, \sigma'_{i+1}) K(\sigma_{i+1}, \sigma_i)}{K^*(\sigma'_{i}, \sigma'_{i+1}) K(\sigma_{i+1}, \sigma_i)}\,
\end{equation}
so that the measurement operator reads
\begin{equation}
   M^{\alpha=0}_{x,i} =
      \frac{\delta_{\sigma_i, -\sigma'_i}
      \delta_{\sigma_{i+1},\sigma'_{i+1}}}{2K^*(\sigma'_{i}, \sigma'_{i+1}) K(\sigma_{i+1},\sigma_i)}.
\end{equation}
In the dissipative case $\alpha\neq 0$ an additional difficulty arises due to the specific path discretization which does not occur for
$P_z$-observables. Namely, the naive ansatz to use $M^{\alpha=0}_{x,i}$ in the expression containing the influence functional is not correct.  To measure $P_x^i$, one also has
to include interactions of $\zeta_i$ (difference paths) with spins at earlier times $j<i$ due to the non-locality in time of the influence functional. It is important to realize that due to the discretization the value of $\zeta_i$ corresponds to the value of the spin path $\zeta(t)$ in the time interval $\left(i-\frac{3}{2} \right)\tau \leq t \leq \left(i - \frac{1}{2} \right)\tau$ so that one has to drop the contribution of the influence functional stemming from the second half of this interval $(i-1)\tau\ldots\left(i-\frac{1}{2} \right)\tau$. To extract expectation values for
$\sigma_x$ and $\sigma_z$ from the {\em same} PIMC sampling, the operator for off-diagonal elements must thus include a 'compensation factor' $\exp\left[\Phi^{\rm off}_i \right]$ with
\begin{eqnarray}
   \Phi^{\rm off}_i & = & \int_0^{\left(i-\frac{1}{2}\right)\tau} dt'
           \zeta_i \Theta \left( t'- (i-1)\tau\right) \nonumber \\
           & \times & \int_0^{t'} dt'' \left[ L'(t'-t'')\zeta(t'') + i L''(t'-t'') \eta(t'') \right] \nonumber \\
           & - & i \mu\int_0^{\left(i-\frac{1}{2}\right)\tau} dt'
           \zeta_i \Theta \left(t'- (i-1)\tau\right) \dot{Q}''(t') \, .
\end{eqnarray}
Here, $\mu$ is due to an initially shifted bath preparation. This correction factor vanishes for diagonal elements ($\zeta_i = 0$). This way, one has for the measurement of
$P_x(u), 0\leq u\leq t$ in the PIMC the operator $ M_{x, i} = M^{\alpha=0}_{x, i} \, \exp(\Phi_i^{\rm off})$
with the discretized form
\begin{eqnarray}
   \Phi^{\rm off}_i & = & \zeta_i \sum_{j=1}^{i} \Lambda^{\rm off}_{i-j} \zeta_j + \zeta_i \sum_{j=1}^{i-1} X^{\rm off}_{i-j} \eta_j
                          \nonumber \\
               &&    + i \zeta_i X_i^{{\rm off},1} \eta_1 + \zeta_i \Lambda_i^{{\rm off},1} \zeta_1
                     + i \zeta_i \hat{X}_i^{(\mu), {\rm off}}\, , \nonumber\\
\end{eqnarray}
where
\begin{eqnarray}
   \Lambda^{\rm off}_0 & = & {\rm Re}\;\left[Q(\tau) - Q(\tau/2) \right] \nonumber \\
   \begin{array}{c} \Lambda^{\rm off}_{1\leq j\leq i-1} \\ X^{od}_j \end{array} & = & \begin{array}{c} {\rm Re} \\ {\rm Im} \end{array}
                  \begin{array}{l} [ Q((j-1/2)\tau) + Q((j+1)\tau) \\ \hspace{6mm} - Q((j+1/2)\tau) + Q(j \tau) ] \end{array} \nonumber \\
   \begin{array}{c} \Lambda^{\rm off,1}_i \\ X^{{\rm off},1}_i \end{array} & = &  \begin{array}{c} {\rm {\rm Re}} \\ {\rm Im} \end{array}
                  \begin{array}{l}  [Q((i-1/2)\tau) - 2Q( (i-1)\tau) \\ \hspace{21mm} + Q((i-3/2)\tau)] \end{array} \nonumber \\
   \hat{X}_i^{(\mu), \rm off} & = & -\nu {\rm Im}\; \left[ Q((i-1/2)\tau) - Q((i-3/2)\tau) \right] \, .\nonumber \\
   \label{Lambda+X_od}
\end{eqnarray}

\end{document}